# Anomalous Features Arising from Random Multifractals

Wei-Xing Zhou, Hai-Feng Liu, and Zun-Hong Yu

East China University of Science and Technology, Box 272

130 Meilong Rd., Shanghai 200237, P.R.China

Fax: 86-21-64250192

Email: wxzhou@moho.ess.ucla.edu

Under the formalism of annealed averaging of the partition function, two types of random multifractal measures with their probability of multipliers satisfying power law distribution and triangular distribution are investigated mathematically. In these two illustrations branching emerges in the curve of generalized dimensions, and more abnormally, negative values of generalized dimensions arise. Therefore, we classify the random multifractal measures into three classes based on the properties of generalized dimensions. Other equivalent classifications are also presented by investigating the location of the zero-point of $\tau(q)$ or the relative position either between the $f(\alpha)$ curve and the diagonal $f(\alpha) = \alpha$ or between the $f(q)$ curve and the $\alpha(q)$ curve. We consequently propose phase diagrams to characterize the classification procedure and distinguish the scaling properties between different classes. The branching phenomenon emerging is due to the extreme value condition and the convergency of the generalized dimensions at point $q = 1$. We conjecture that the branching condition exists and that the classification is universal for any random multifractals. Moreover, the asymptotic behaviors of the scaling properties are studied. We apply the cascade processes studied in this paper to characterizing two stochastic processes, i.e., the energy dissipation field in fully developed turbulence and the droplet breakup in atomization. The agreement between the proposed model and experiments are remarkable.

# 1 Introduction

In the present formalism of either the restricted theory of multifractal[1-5] or the general theory of multifractal[6-10], multifractal measures are decomposed into interwoven fractal sets each of which is characterized by its singularity. In the deterministic case, the singularity spectrum $f(\alpha)$ is always non-negative and varies in a finite range[11,12]. However, when one investigates the multifractal nature of random measures arising from experiments, in particular the diffusion-limited aggregation[13] and the dissipation field of turbulence[14], negative dimensions are discovered, which means that $f(\alpha)$ can be negative for certain $\alpha$. The earliest theoretical insight into negative dimensions owed to Mandelbrot[15], which should date back to 1974 and also see Ref. [16]. Further studies were mostly carried out by Mandelbrot[17-19]. To calculate the negative part of the $f(\alpha)$ function, Chhabra & Sreenivasan proposed an efficient procedure named multiplier method that can extract the $f(\alpha)$ spectrum with exponential less work and is more accurate than the conventional box-counting method[20-23].

As pointed out by Mandelbrot and his coworkers [6,7,10], there are two meanings of the term "multifractal". The earlier and more general meaning comes from the notation of "multiplicative cascade that generates nonrandom or random measures", and describes "measures that are multiplicatively generated". This meaning results in the general theory of multifractal (GTMF). A second meaning describes "a nonrandom measure for which it is true for all $-\infty < q < +\infty$ that the partition function scales like a power of the form of $\varepsilon^{\tau(q)}$", and is referred to as the restricted theory of multifractal (RTMF). RTMF is a theory corresponding to deterministic multiplicative cascade process, while GTMF corresponding to both deterministic and random multiplicative cascade process.

In GTMF, two features, the single-sided multifractals[6-10] and negative dimensions[14,17-20], have been discovered when comparing with RTMF. The appearance of left-sided multifractals is stimulated directly by the discovery of phase transform in DLA[24-27], although the idea was hidden in Mandelbrot's 1974 measure. In the framework of left-sided multifractals, the definition domain of $q$ is $[0,+\infty)$, while that is $(-\infty,1]$ in the case of right-sided multifractals[7,9]. Furthermore, the singularity strength $\alpha$ may tend to infinity when $q$ tends zero for left-sided multifractal measures. Similar conclusion can be drawn to the right-sided multifractal measures because of the symmetry between left- and right-sided multifractals. Another feature is the existence of negative dimensions, which is expected to be more universal than single-sidedness. The negative dimensions arise from either the intrinsic randomness or a random view of a deterministic process, say experiments in turbulence. A simple but cogitative example to account for this point is referred to the random binary process[20]. However, in the discrete case of random multifractals, the existence of negative dimensions is not indispensable[28]. The formalist corresponding to negative dimensions in random multifractal measures relates to Cramer's theorem of large deviations[18,29].

A random multiplicative cascade process will generate a random multifractal measure on certain geometry support, which is a stochastic object. Two averagings are valid when investigating the scaling properties of such stochastic objects. One can choose to define either an annealed scaling exponent or a quenched one, respectively. Halsey[30-31] has expected the quenched averaging to yield a more physical result in DLA. However, the quenched averaging cuts off the intrinsic or practically deduced randomness for many processes, say the fully developed turbulence, where the annealed averaging shows its advantage in characterizing the lacunarity of certain occasionally emerging measures[20,32].

First, perform a random multiplicative cascade process. Divide uniformly the interval $[0,1]$ into $b$ pieces with the multipliers picked randomly from $[M_{i1},\cdots,M_{ib}]$ of sizes $1/b$ with the probability $p_i$, where the subscript $i=1,2,\cdots,K$ and $\sum_{i=1}^{K} p_i = 1$. At the next generation each piece of the set is further divided into $b$ pieces, each with a randomly picked generator. This procedure is continued *ad* infinity. It is clear that the multiplicative process must produce a multifractal measure. More generally, one can investigate a random multifractal measure rearranged by multipliers with continuous probability density $\Pr(M)$.

From Cramer's theorem of large deviations[33-34], the mass exponent can be defined by an annealed averaging of moments of the multipliers, namely

$$\tau(q) = -D_0 - \frac{\log\langle M^q \rangle}{\log b}, \tag{1}$$

and the multifractal spectrum $f(\alpha)$ is linked with $\tau(q)$ by Legendre transform and inverse Legendre transform[32]. Therefore

$$\alpha(q) = \tau'(q) = -\frac{\langle M^q \log M \rangle}{\langle M^q \rangle \log b}, \tag{2}$$

and

$$f(\alpha(q)) = q\alpha(q) - \tau(q) = \frac{\langle M^q \rangle \log\langle M^q \rangle - \langle M^q \log M^q \rangle}{\langle M^q \rangle \log b} + D_0. \tag{3}$$

Similar to that in Ref. [35], Eq. (3) can be rewritten in the form

$$f(\alpha(q)) = D_0 - \frac{1}{\log b} \cdot \left\langle \left( \frac{M^q}{\langle M^q \rangle} \right) \log\left( \frac{M^q}{\langle M^q \rangle} \right) \right\rangle. \tag{4}$$

We have already discussed the continuous situation with the multipliers exponentially distributed and "anomalous" feature were discovered[36]. In this paper, we investigate the random multifractals with other two types of continuously distributed multipliers and similar new features are discovered.

## 2 Mathematical Illustrations

### 2.1 Power law distribution

Now, consider an example in which the probability density of the multipliers is in the form

$$\Pr(M) = (x+1)M^x, \tag{5}$$

where the real number $x \neq -1$ is a parameter. The annealed averaged moments is

$$\langle M^q \rangle = \frac{x+1}{q+x+1}. \tag{6}$$

It is obvious that $q > -x-1$ and $x > -1$. As we can see, the definition domain is somewhat similar to that of single-sided multifractals. Simple derivations result in as follows.

$$\tau(q) = -1 - \log_b\left(\frac{x+1}{q+x+1}\right). \tag{7}$$

$$D_q = -\frac{1}{q-1}\left[\log_b\left(\frac{x+1}{q+x+1}\right)+1\right]. \tag{8}$$

$$\alpha(q) = \frac{1}{(q+x+1)\log b}. \tag{9}$$

$$f(q) = \log_b\left(\frac{x+1}{q+x+1}\right) + \frac{q}{(q+x+1)\log b} + 1. \tag{10}$$

We have found that, there are three different classes of the generalized dimensions corresponding to distinct parameter range. The cause, which cannot be given here, will be discussed in details later. These classes are followed.

Class I. For $x > (2-b)/(b-1)$, there are two separated branches in the curve of generalized dimensions, each with an extreme point. The definition domain is the union of sets $\{-x-1 < q < 1\}$ and $\{q > 1\}$, which is discontinuous. On the left branch, $D_q$ tends to $+\infty$ when $q$ traces near to the left-most boundary or right boundary $q = 1^-$. On the right branch, $D_q$ tends to $-\infty$ when $q$ tends towards the left boundary, and to $0^+$ when $q$ tends to infinite. Shown in Figs. 1-2 are the typical pictures with $x=3$ and $b=2$, and the two extreme points as well.

Class II. For $x=(2-b)/(b-1)$, there is only one continuous branch, which is located in the first quadrant, without any extreme point. The definition domain is $\{q>-1\}$. Note that the curve is continuous at $q=1$ in the sense of limit, which is the only case in which the generalized dimension converges to a finite value, $2/\log b$. The chart is shown in Fig. 3 with $b=2$.

Class III. For $-1<x<(2-b)/(b-1)$, there are two separated branches without any extreme points in the curve. The definition domain is the union of sets $\{-x-1<q<1\}$ and $q>1$, which is discontinuous at point $q=1$. The two branches are both divergent at the point $q=1$. Figures 4-5 are the typical pictures with $x=-0.5$ and $b=2$. Note that, when $x\to(2-b)/(b-1)$, the two branches tend to the curve of $x=(2-b)/(b-1)$, and the divergences near $q=1$ become more and more slowly. Therefore, to detect the disparity between the two cases, one should be more careful and increase the computation spacing and accuracy. This phenomenon can be viewed as a break from disconnectedness to connectedness.

Furthermore, we can obtain the phase diagram of components $b$ and $x$ to determine that to which class the fixed pair $(b,x)$ belongs. Shown in Fig. 6 is the phase diagram. The overall phase space is $\{(b,x):b>1,x>-1\}$. The dashed axes are excluded. One can see that, the region corresponding to Class II is a continuous curve, which is the boundary between Class I and III. The curve can obviously be regarded as *critical curve*.

The typical diagrams of $\tau(q)$, $\alpha(q)$, $f(q)$ and $f(\alpha)$ with $b=2$ and $x=-0.5$, $x=0$ and $x=3$ are illustrated in Figs. 7-9, respectively. In these figures, the solid lines indicate $x=-0.5$, while the dashed and dotted lines indicate respectively $x=0$ and $x=3$.

Unlike the generalized dimensions where the type of the curves depends on the parameter $x$,

all the curves with diverse $x$ are similar and show similar asymptotic behaviors. The asymptotic behaviors of $D_q$, $\tau(q)$, $\alpha(q)$ and $f(\alpha)$ when $q$ tends to $-x-1$ and/or infinite are listed in Table 1, which are derived directly from Eqs. (7)-(10) and can be seen in Figs. 1-5 and Figs. 7-9 as well. For the generalized dimensions, the asymptotic behaviors when $q \to 1^+$ or $q \to 1^-$ are discussed previously and can also be see in Figs. 1-5.

Eliminating $q$ in Eqs. (9)-(10), we obtain the explicit expression of the singularity spectrum versus singularity strength:

$$f(\alpha) = \log_b[\alpha(x+1)\log b] - (x+1)\alpha + 1 - 1/\log b . \qquad (11)$$

Therefore,

$$f'(\alpha) = \frac{1}{\alpha \log b} - (x+1) . \qquad (12)$$

If $\alpha \to +\infty$, $f'(\alpha) \approx -(x+1) < 0$, which implies that $f(\alpha)$ decreases at a constant rate. On the other hand, if $\alpha \to 0^+$, $f'(\alpha) \approx 1/(\alpha \log b) > 0$. Hence, $f(\alpha)$ decays more quickly when tending near to 0 than to infinite. In addition, $f(\alpha(1)) \neq \alpha(1)$ except for $x = 1/(b-1) - 1$. Therefore, the multifractal spectrum need not be tangent to the diagonal of the first quadrant any longer, which is a universal property for deterministic multifractals and breaks down for the random multifractal measures.

## 2.2 Triangular distribution

Our second example is to get a further insight into a class of random measures generated by a cascade process whose multipliers are chosen randomly from a triangular distribution, namely,

$$\Pr(M) = \begin{cases} 2M/x, & 0 < M < x, \\ 2(M-1)/(x-1), & x < M < 1, \end{cases} \qquad (13)$$

where $x \in (0,1)$ is a parameter. If $x = 0.5$, it reduces to the first example shown in Ref. [20]. We have

$$\langle M^q \rangle = \int_0^1 M^q \Pr(M) dM = \frac{2(1-x^{q+1})}{(q+1)(q+2)(1-x)} . \qquad (14)$$

According to the definition of $\tau(q)$, $\langle M^q \rangle$ must be positive. Hence, we have $q > -2$. Therefore,

$$\tau(q) = \log_b \frac{(q+1)(q+2)(1-x)}{2(1-x^{q+1})} - 1, \tag{15}$$

$$D_q = \frac{1}{q-1}\left[\log_b \frac{(q+1)(q+2)(1-x)}{2(1-x^{q+1})} - 1\right], \tag{16}$$

$$\alpha(q) = \frac{2q+3}{(q+1)(q+2)\log b} + \frac{x^{q+1}\log x}{(1-x^{q+1})\log b}. \tag{17}$$

Let $q \to -1$, then $\tau \to -1 - \log_b \frac{2\log x}{x-1}$ and $\alpha \to \frac{2-\log x}{2\log b}$, which implies that $D_q$ and $f(q)$ are continuous at $q = -1$ as well. To make $D_q$ converges when $q \to 1$, the sufficient and necessary condition is $x = (3-b)/b$, implying $1.5 < b < 3$, which is again the critical line between two different types of generalized dimensions. Three distinct classes arise as follows.

Class I. For $(3-b)/b < x < 1$, there are two separated branches in the curve of generalized dimensions, each with an extreme point. The two extreme values correspond to the solutions of the nonlinear equation $D'_q = 0$.

Class II. For $x = (3-b)/b$, there is a single continuous curve with no extreme value.

Class III. For $0 < x < (3-b)/b$, there are two separated branches with no extreme value.

The typical diagrams of the three cases are illustrated in Figs. 10-12, respectively. The shape of the left branch in Fig. 10 is similar to the curve in Fig. 1, while the right one is similar to that in Fig. 2. The shape of the curve in Fig. 11 is similar to Fig. 3. And that, the shape of the left branch in Fig. 12 is similar to the curve in Fig. 4, while the right one is similar to that in Fig. 5. Meanwhile, the asymptotic behaviors of the generalized dimensions in the present distribution are similar to those in the power law distribution. The phase diagram of the components $b$ and $x$ is shown in Fig. 13. The phase space is $\{(b,x): b > 1, 0 \leq x \leq 1\}$. Note that, $x = 1$ and $x = 0$ are included as well, which will be clear below.

For the sake of completeness, consider two cases of $x = 1$ and $x = 0$. If $x$ is equal to unity, the probability distribution of multipliers is $\Pr(M) = 2M$, where $0 < M < 1$. Therefore, one finds that it is the case argued in the previous section.

When $x = 0$, we have $\Pr(M) = 2(1-M)$, where $0 < M < 1$, which leads to $\langle M^q \rangle = 1/(q+1)(q+2)$. Henceforth, the definition domain is $\{q : q > -1 \quad or \quad q < -2\}$, which is different to those corresponding to $0 < x \leq 1$. When $q > -1$, the critical point is $b = 3$, and the corresponding curve of $D_q$ is similar to that in Fig. 11. On the same time, the diagrams of $D_q$ with $1 < b < 3$ and $b > 3$ are similar to those shown in Figs. 10 and 12, respectively. When $q < -2$, the curve of $D_q$ is similar to the origin-symmetric plot of that illustrated in Fig. 2. These three cases are illustrated in Figs. 14-16.

The curves of mass exponents, singularity strengths and singularity spectra in the overall definition domain with the bases $2$, $3$ and $4$ are shown respectively in Figs. 17-19. Since there are two disconnected parts of the definition domain, all these curves are branched. However, are the left branches physically meaningful? In other words, how can a Holder exponent be negative? An interesting explanation is the death or survival criterion. Following Mandelbrot[19], the limit measure generated by triangular multipliers with $x = 0$ is identically zero, which is identified by the "microscope" of $q < -2$, since the probability density of the multiplier near $M = 0$ is positive, while that in the other cases of $0 < x \leq 1$ is zero. Therefore, any characteristic quantities corresponding to $a < 0$ are meaningless and die, while those quantities corresponding to $\alpha > 0$ survive, and so do the measures as well. Consequently, the $x = 0$ case is involved in the previous three classes. Hence, the parts with $q < -2$ are nothing meaningful physically, and the definition domain in the present case is identical to those of $0 < x \leq 1$.

Now, go back to the phase diagram. We can draw a conclusion that, there are only three classes when $0 \leq x \leq 1$ with the great divide $x = (3-b)/b$. Similar curves will be obtained in certain class, and those in different classes are dissimilar to each other.

## 3 Discussion

### 3.1 Branching condition and extreme value condition

In deterministic multifractals, such as RTMF, single-sided multifractals, and discrete random multifractals, one find than $f(\alpha) \leq \alpha$, and that the equation $f(\alpha) = \alpha$ has a single root $\alpha(1)$. We have $f - \alpha \sim O((q-1)^2)$ when $q \to 0$, since

$$D'_q = (f - \alpha)/(q-1)^2 \tag{18}$$

In these cases, no branches appear. However, in the continuous random multifractals, things go in a fairly different way, just as the two cases in Sec. 2. The appearance of branching of the generalized dimensions is caused by the divergence of $D_q$ when $q \to 1$. In the non-branching cases, $D_1$ is a finite value, which means that $\tau(q) \sim O(q-1)$ when $q \to 1$. Considering the two cases in Sec. 2, there is a function $x = h(b)$, which leads to non-branching of $D_q$. And also the $f(\alpha)$ curve is tangent to the linear line $f(\alpha) = \alpha$ at $q = 1$. Therefore, we regard $x \neq h(b)$ as the branching condition (BC), such measures are non-conservative.

When $x > h(b)$, the extreme value condition (EVC) is satisfied. There are roots of the equation $D'_q = 0$, where extreme values are reached in the curve of $D_q$. In this case, there exist $q_1$ and $q_2$ satisfying $D'_{q_1} = 0$ and $D'_{q_2} = 0$ where $q_{bottom} < q_1 < 1 < q_2$. Therefore we have $D'_q > 0$ for $q_1 < q < 1$ and $1 < q < q_2$, and $D'_q < 0$ for $q_{bottom} < q < q_1$ and $q > q_2$. Consequently, there are two roots $\alpha_1$ and $\alpha_2$ of the equation $f(a) = \alpha$ implying the intersecting between the $f(\alpha)$ curve and the linear line $f(\alpha) = \alpha$.

When $x < h(b)$, the EVC fails. There is no root of the equation $D'_q = 0$ and consequently no extreme point arises in the curve of $D_q$. In this case, $D'_q < 0$ for $q_{bottom} < q < 1$ and $q > 1$. Henceforth, $f(\alpha) < \alpha$ for all $q$. The $f(\alpha)$ curve is separated to the linear line $f(\alpha) = \alpha$.

Moreover, the definition domain of $D_q$ excludes the point $q = 1$, while that of $\tau(q)$, $\alpha(q)$ and $f(q)$ is connected at $q = 1$.

### 3.2 Classification of random multifractals

Following the discussion in the previous subsection, we can classify random multifractal measures into three classes according to the relative position between the $f(\alpha)$ curve and the diagonal $f(\alpha) = \alpha$ of the first quadrant.

Class I: Intersection. If the diagonal $f(\alpha) = \alpha$ intersects the $f(\alpha)$ curve, there must exist two intersecting points $(\alpha(q_1), f(q_1))$ and $(\alpha(q_2), f(q_2))$. Without loss of universality, we can regard that $q_1 < 1 < q_2$, since it is impossible that they intersect at $q = 1$ or the same side of $q = 1$. Therefore, the solution set of Eq. (18) is $\{q_1, q_2\}$, which correspond to the two extreme points of the $D_q$ curve. The existence of extreme points is the sign of branching in the $D_q$ curve.

Class II: Tangency. This is the only case that one can use the so-called determination criterion[37] to judge whether the computed $f(\alpha)$ of random multifractal measures is valid or not. In this case we have $f(\alpha) \leq \alpha$, where the equality holds when $\alpha = \alpha(1)$. We conjecture, in the more general sense, that no branching appears in the curve of generalized dimensions, and that no extreme points exist as well.

Class III: Separation. In this case, the $f(\alpha)$ curve locates below the diagonal line $f(\alpha) = \alpha$, implying $f(\alpha) < \alpha$ for all $q$ and hence $\alpha$ as well. We also conjecture, in a more general sense, that the branching in the curve of generalized dimensions emerges again, and that no extreme points exist.

We expect that the conjectures are universal for measures generated from random multiplicative cascade process with its multipliers picked from certain continuous probability distribution. Nevertheless, these conjectures still need further verifications or a rigorous proof. Note that the classification does nothing with these conjectures and is universal.

An alternative way is to investigate the relative position between the curves of $f(q)$ and $\alpha(q)$ with intersection, tangency and separation. The forthcoming subsection will show another equivalent classification method via analyzing $\tau(q)$. Therefore, three equivalent rules are established to classify random multifractal measures, which come from the natures of $D_q$, $\tau(q)$ and $f(\alpha)$, respectively. It seems that one can't intuitively classify such measures via investigating the properties of $f(q)$ or $\alpha(q)$. The reason is because that, each of the $D_q$, $\tau(q)$ and $f(\alpha)$ can characterize fully the multifractal measures, and that they can transform from each one to others and are consequently equivalent to each other.

### 3.3 Negative generalized dimensions

In the previous section, negative generalized dimensions were discovered in Class I and III. It seems to be anomalous and is possibly a new feature. The cause of the appearance of negative generalized dimensions can be found when investigating the definition of $D_q$ from $\tau(q)$. Assume $q_0$ satisfies $\tau(q_0) = 0$. Note that $\tau'(q) > 0$. If $q_0 > 1$, $\tau(q) < 0$ for any $1 < q < q_0$, and hence $D_q$ is negative. If $q_0 = 1$, $D_q$ is positive for all $q$ in the definition domain. If $q_0 < 1$, $\tau(q) > 0$ for any $q_0 < q < 1$, and hence $D_q$ is negative. An intuitive view is demonstrated in Figs. 7 and 17. It is obvious that one can classify continuously random multifractal measures according to the value of $q_0$, which is equivalent to the classification method presented previously.

### 3.4 Asymptotic behaviors of generalized dimensions and singularity strengths

Generally, the tendencies of $D(q)$ and $\alpha(q)$ are similar to each other. Moreover, both the minimal and maximal values of $D(q)$ and $\alpha(q)$ exist and are identical respectively[11,12] in RTMF, while in left-sided multifractal measures[6], the minimal $\alpha(q)$ exists and $\alpha \to +\infty$ when $q \to +\infty$. However, there are no boundaries for $\alpha(q)$ in its definition domain for the present case. As shown in Figs. 8 and 18, $\alpha \in (0, +\infty)$. Note that, $\alpha(q)$ cannot reach 0 as its minimal. As pointed out by Mandelbrot[6], a sufficient condition for $\alpha_{min} = 0$ is that one can identify at least one point where $\alpha = 0$. To meet $\alpha = 0$, one should investigate the maximal measure 1, which can never be reached since all multipliers are less than unique although one can approach it as near as possible. Therefore, it is universal that, in the case of discrete probability distribution of the multipliers, $\alpha_{min}$ exists which corresponds to the region with maximal measure[20], while it is not the matter in the continuous case.

### 3.5. Non-integer base and phase diagram

Formally, we suggest that non-integer base is also valid when investigating such random cascade process that the numbers of the rules are not identical, where the non-integer base can be looked upon as an averaged base $\langle b \rangle$. If one considers only the integer bases, the phase space is a set of discrete points.

## 4 Applications

### 4.1 A random multiplicative cascade model for fully developed turbulence

The so-called $p$-model is based on the deterministic binomial multiplicative cascade, with two multipliers $M_1 = 0.3$ and $M_2 = 0.7$ of scale ratios $l_1 = l_2 = 0.5$. The results are in remarkable agreement with experimental ones[38]. Nevertheless, negative part of the multifractal spectrum was neglected, which should appear in turbulence[14,20,39], since no randomness was considered in the $p$-model. To specify the negative part of $f(\alpha)$, randomness must be considered.

Consider a random multiplicative cascade model, in which a power-law multiplier distribution with $b = 1.3$ and $x = 7/3$ is adopted, namely, $\Pr(M) = 10/3 M^{7/3}$. The analytical expression of $f(\alpha)$ can be obtained according to Eq. (11). The resultant multifractal spectrum is illustrated in Fig. 20 represented by the solid line. The dashed line represents the multifractal spectrum obtained from the $p$-model[33], while the joined circles denote the spectrum from the atmospheric surface layer using 720000 data points according to the multiplier method[20]. Obviously, the agreement within the three plots is excellent when $q > 0$.

Since we cannot find the raw experimental data, the results of the experiment is from Ref. [20], and the right part of $f(\alpha)$ has to be neglected. Nevertheless, it is expected to fit the random multifractal model perfectly well. When considering randomness in turbulence, the $p$-model is not suitable for the description of turbulent energy dissipation any more. The right parts of the two models have shown great difference. In addition, the tendencies near the domain of $f = 0$ are remarkably distinct.

Another difference between the two models is the value of the base. In the $p$-model, the small scales from large ones are space-filling. Meanwhile, in the present random multifractal model, the small scales are less and less space-filling since $b = 1.3$. In this sense, the random multifractal model is somewhat similar to the $\beta$-model[40-41]. That the base $b$ less than 2 implies that, only a portion of the generated eddies are active, and $b = 1.3$ can be regarded as an expectation of the overall cascade process.

## 4.2 Random multiplicative model for drop breakup in atomization

To characterize the droplet size distribution in the atomization process, a random cascade model has been presented[42]. The model fits the experimental results very well from the fractal point of view. This model can be re-described as follows. The breakup of droplet follows a random cascade process. Each piece at a fixed generation may either splits into two smaller pieces or retains one. The splitting probability is denoted by $p$. The probability distribution of the mass ratio $M$, or multiplier, is uniform. Therefore, we have

$$\Pr(M) = \begin{cases} p, & 0 < M < 1 \\ (1-p)/2, & M = 0, 1 \end{cases} \qquad (19)$$

Postulating that the droplets atomized by the high-speed jet are spatially uniformly dispersed in the spray zone because of strong plume. Then the annealed average of $M$ is given by

$$\langle M^q \rangle = \frac{1-p}{2} + \int_0^1 pM^q dM = \frac{1-p}{2} + \frac{p}{q+1}. \qquad (20)$$

Therefore, the multifractal spectrum can be obtained theoretically. The solid line shown in Fig. 21 is the singularity spectrum from the present model with $p = 1$. The measurement of the droplets size distribution is carried out using Dual Particle Dynamic Analyzer (Dual PDA). A record with 525287 data points is obtained and then analyzed applying the multiplier method to get the multifractal spectrum, which is illustrated in Fig. 21 with solid dots.

The agreement between the experiment and the model is perfectly good except the right tail of the curves. The cause is from the breakup mechanism of droplets. We have estimated the Weber numbers of droplets throughout the spray region according to the experiments. The maximum of Weber number is less that $21.5$. A majority of droplets in the spray had the Weber numbers less that the critical value of $12$, indicating that these drops lie in the vibrational breakup regime[43-44]. Meanwhile, the rest droplets had the Weber number between 12 and 21.5 falling in the bag breakup regime. In the vibrational breakup regime, one droplet splits into two sub-droplets with the mass ratio of sub-droplet to its mother droplet around 0.5, while in the bag breakup regime, one droplet splits into several relatively bigger sub-droplets and many smaller sub-droplets. Therefore, vibrational breakup dominates and bag breakup will also arise. In the case of bag breakup, we can regard the mother droplet as several dummy droplets. However, in the tail of the right part of the multifractal spectrum, the decay of the experimental curve is much faster than that from the model. Hence, the difference between the two regimes increases with decreasing negative $q$. This may be a universal property when comparing multifractal spectra arising from continuous and discrete multiplier probability distribution, since, in general, $0 < \alpha < \alpha_{\max}$. The $f(\alpha)$ curve is tangent to linear line $f(\alpha) = \alpha$ when the multiplier method is used, as shown in Figs. 20-21. A more detailed argument can be found in Ref. [44].

## 5 CONCLUSIONS

In this paper, we enriched this deep idea of Mandelbrot, which was presented in Ref. [27], with two mathematical illustrations. For non-conservative random multifractals with continuously distributed multipliers, the curve of generalized dimensions branches, and more abnormally, negative values of generalized dimensions arise. We classified the random multifractal measures into three classes based on the properties of generalized dimensions. We found that, one can perform the equivalent classification by investigating the location of the zero-point of $\tau(q)$ or the relative position either between the $f(\alpha)$ curve and the diagonal $f(\alpha) = \alpha$ or between the $f(q)$ curve and the $\alpha(q)$ curve. Consequently, we presented a phase diagram to characterize the classification procedure and distinguish the scaling properties between different classes. For fixed base $b$ or parameter $x$, phase transition with crossover from Class I to Class III appears. We except most of these properties are universal for random multifractals.

The detection of branching phenomenon emerging in the curve of the generalized dimensions follows a two-step procedure. If the extreme value condition fits, the investigated measure belongs to Class I. Otherwise, if the generalized dimensions converge at point $q=1$, the measure lies in Class II. Absence both of EVC and convergence at $q=1$ indicates the measure to fall in Class III. We also conjectured that the branching condition exists in a set of random multifractal measures.

Furthermore, we studied two stochastic processes about the modeling of the energy dissipation field in fully developed turbulence and the droplet breakup in atomization. The random multiplicative cascade models presented can characterize these processes perfectly well.

**Acknowledgements**

This research was supported by the National Development Programming of Key and Fundamental Researches of China (No. G1999022103).

TABLE I. Asymptotic behaviors

|         | $q \downarrow -x-1$ | $q \uparrow +\infty$ |
|---------|:-------------------:|:--------------------:|
| $D_q$   | $+\infty$           | $0^+$                |
| $\tau(q)$ | $-\infty$         | $+\infty$            |
| $\alpha(q)$ | $+\infty$       | $0^+$                |
| $f(\alpha)$ | $-\infty$       | $-\infty$            |

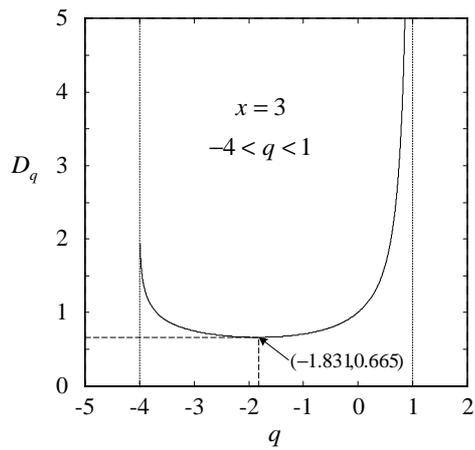

Figure 1. Typical chart of the left branch of the first class with $x = 3$ and $b = 2$.

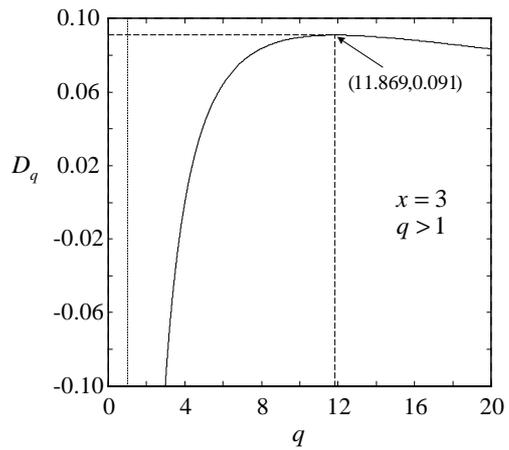

Figure 2. Typical chart of the right branch of the first class with $x = 3$ and $b = 2$.

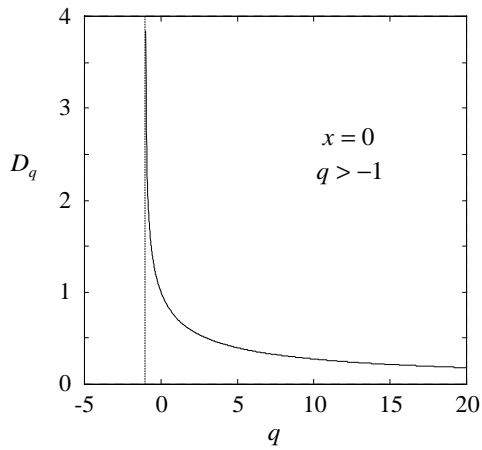

Figure 3. Typical chart of the second class with $x = 0$ and $b = 2$.

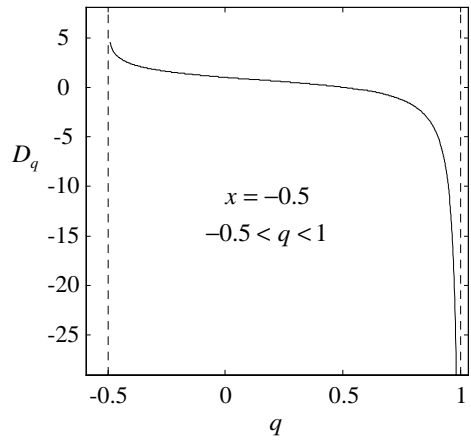

Figure 4. Typical chart of the left branch of the third class with $x = -0.5$ and $b = 2$.

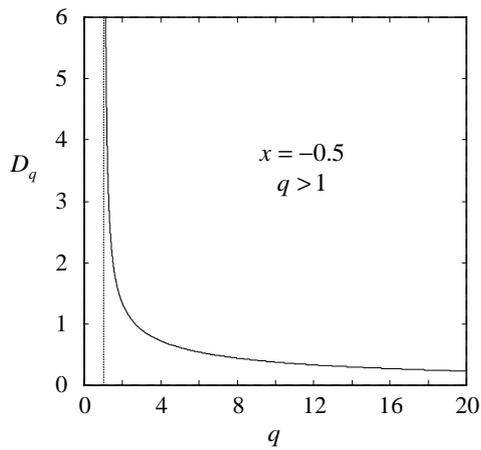

Figure 5. Typical chart of the right branch of the third class with $x = -0.5$ and $b = 2$.

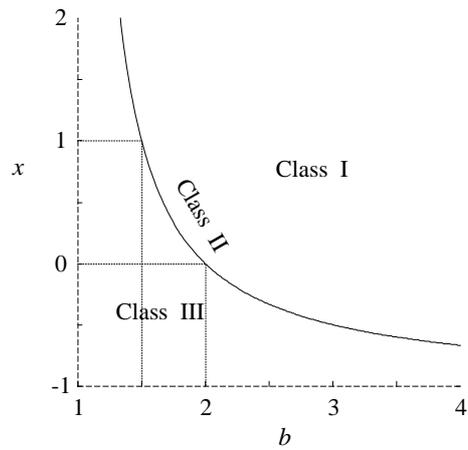

Figure 6. Phase diagram of the power distribution.

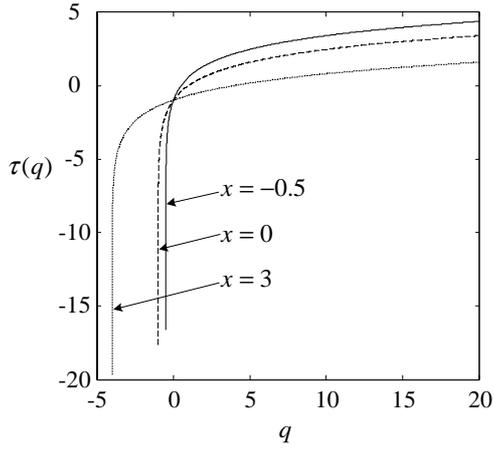
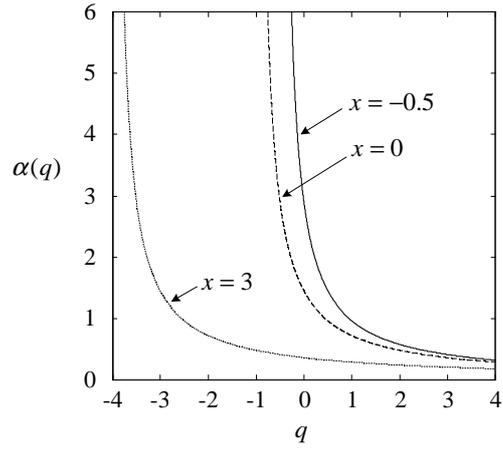

Figure 7. Typical diagrams of mass exponent.    Figure 8. Typical diagrams of singularity strength.

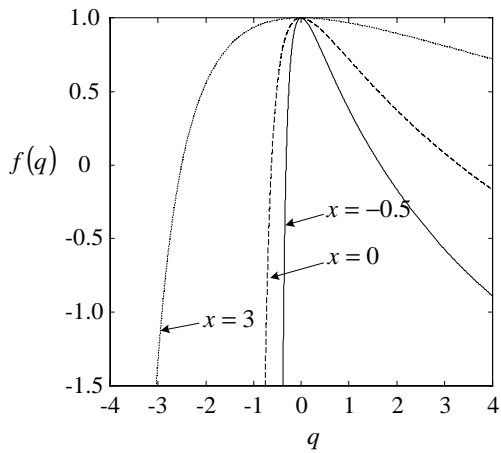
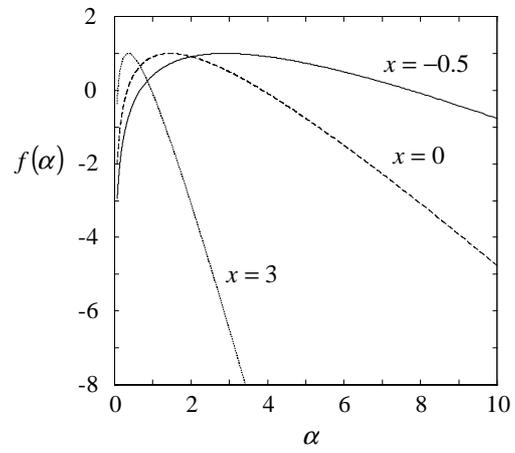

Figure 9. Typical diagrams of multifractal spectra

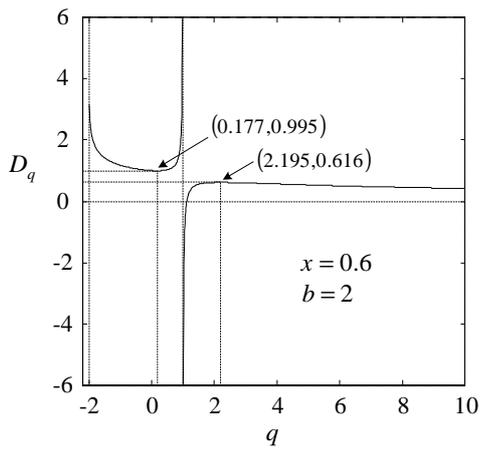
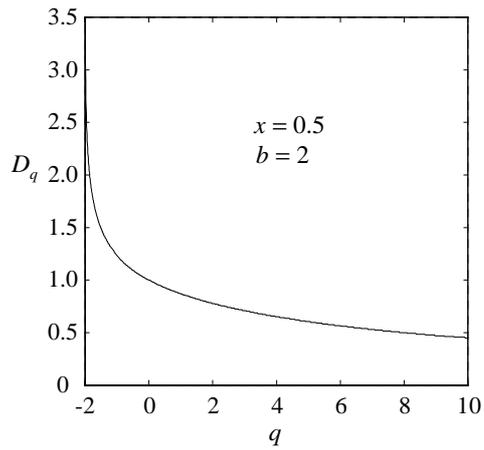

Figure 10. Typical diagram of the first class    Figure 11. Typical diagram of the second class
with $x = 0.6$ and $b = 2$.                      with $x = 0.5$ and $b = 2$.

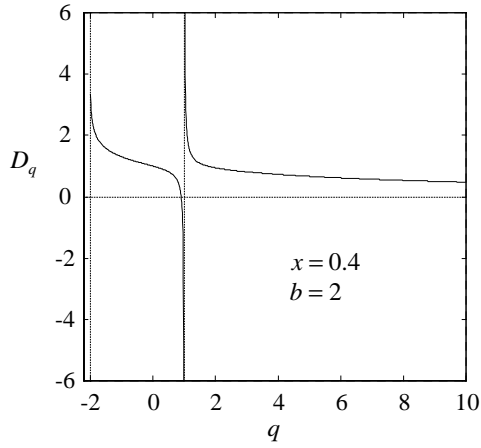

Figure 12. Typical diagram of the third class with $x = 0.4$ and $b = 2$.

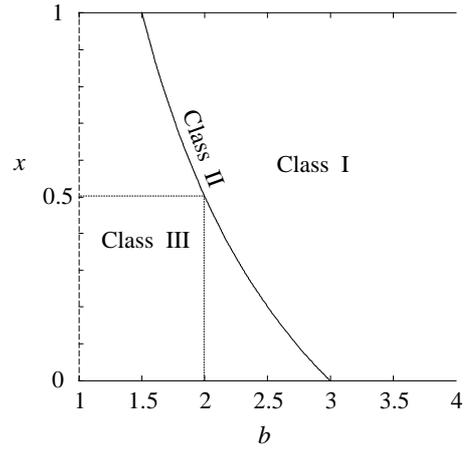

Figure 13. Phase diagram of the triangular distribution.

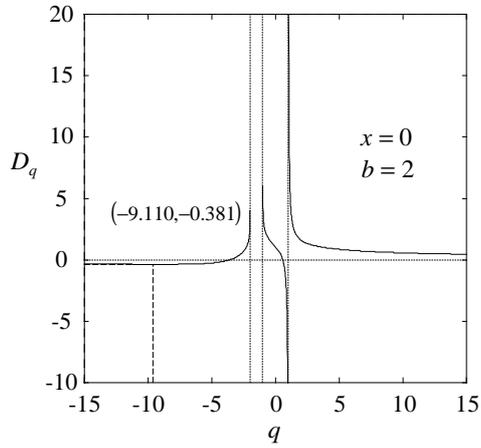

Figure 14. The generalized dimensions in the case of $x = 0$ and $b = 2$.

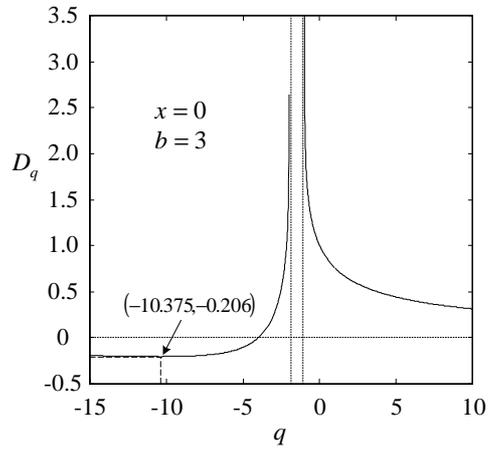

Figure 15. The generalized dimensions in the case of $x = 0$ and $b = 3$.

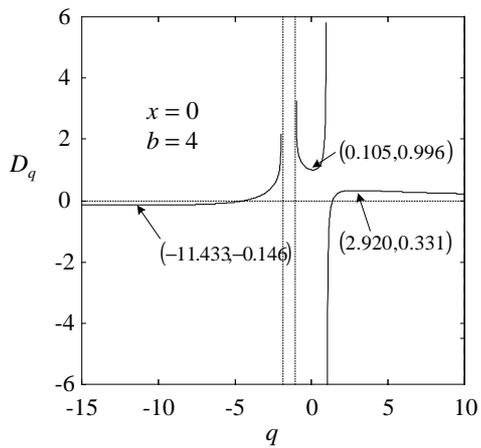

Figure 16. The generalized dimensions in the case of $x = 0$ and $b = 4$.

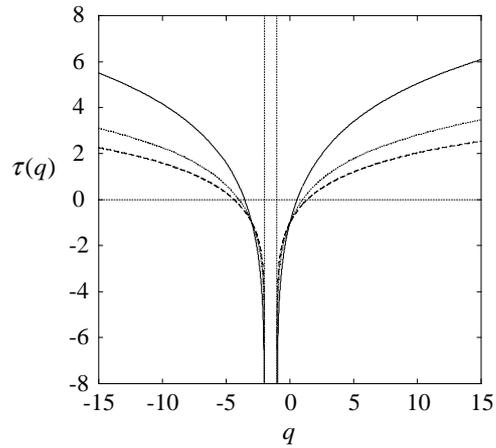

Figure 17. Diagrams of $\alpha$ in the classes of $x = 0$, and $b = 2$ (dashed), 3(dotted) and 4(solid).

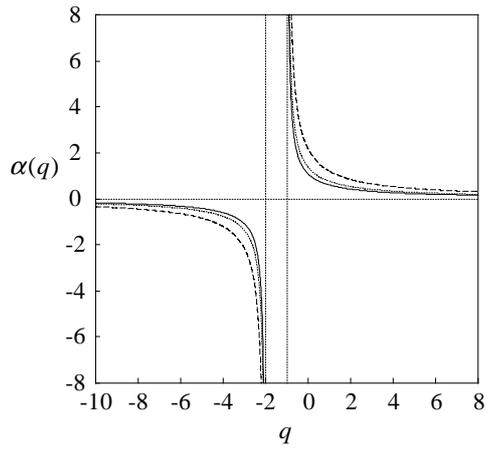

Figure 18. Diagrams of $\alpha$ in the classes of $x = 0$, and $b = 2$ (dashed), 3(dotted) and 4(solid).

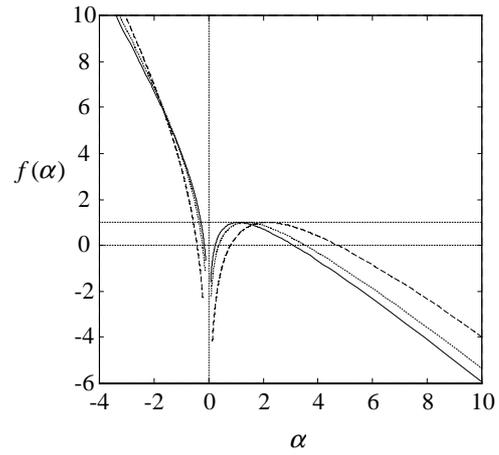

Figure 19. $f$ versus $\alpha$ in the classes of $x = 0$, and $b = 2$ (dashed), 3(dotted) and 4(solid).

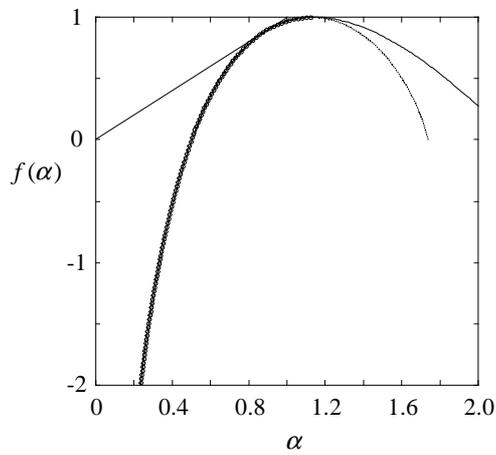

Figure 20. Comparison within p-model (dotted), the multiplier method (joined circles) and the random multifratal model (solid).

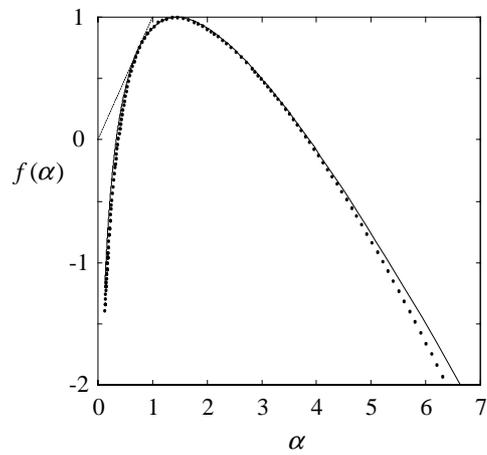

Figure 21. Comparison between the experiment result (solid dots) and the random multifratal model (solid line) of droplet breakup.

23